\documentclass[runningheads]{llncs}

\usepackage{graphicx}
\usepackage{amsmath}
\usepackage{amssymb}
\usepackage{hyperref}
\usepackage{color}

\usepackage[ruled,vlined,linesnumbered]{algorithm2e}
\usepackage{algpseudocode}
\usepackage{booktabs}
\usepackage{todonotes}
\usepackage{mathtools}
\usepackage{units}
\usepackage{comment}
\usepackage{marvosym}

\newcommand{\T}{\top}

\newcommand{\e}[2]{\ensuremath{ \text{#1e#2} }}

\iffalse
\newcommand{\vw}[1]{}%{\todo[color=red]{\emph{VW}: #1}}
\newcommand{\luca}[1]{}%{\todo[color=green]{\emph{LB}: #1}}
\newcommand{\MB}[1]{}
\newcommand{\GG}[1]{}
\renewcommand{\todo}[1]{}
\else
\newcommand{\vw}[1]{\todo{\emph{VW}: #1}}
\newcommand{\luca}[1]{\todo{\emph{LB}: #1}}
\newcommand{\GG}[1]{\todo{\emph{GG}: #1}}
\newcommand{\MB}[1]{\todo{\emph{MB}: #1}}
\fi

\spnewtheorem{model}[theorem]{Model}{\bfseries}{\itshape }

\begin{document}
%\title{Building Bridges
%in Markovian\\ Population Models using Lumping}
% \title{Solving the Bridging Problem\\in Markovian Population Models\\using State-Space Lumping}
\title{Analysis of  Markov Jump Processes under Terminal Constraints}
\author{Michael Backenköhler\inst{1}\textsuperscript{,\Letter}, Luca Bortolussi\inst{2,3}, Gerrit Großmann\inst{1}, Verena Wolf\inst{1,3}}
\authorrunning{M.\ Backenköhler et al.}
\institute{
${}^1$Saarbrücken Graduate School of Computer Science,
Saarland University,
Saarland Informatics Campus E1 3,
Saarbrücken, Germany\\
\textsuperscript{\Letter} \email{michael.backenkoehler@uni-saarland.de}\\
${}^2$Univeristy of Trieste, Trieste, Italy\\
${}^3$
Saarland University,
Saarland Informatics Campus E1 3,
Saarbrücken, Germany\\
% \email{michael.backenkoehler@uni-saarland.de}\\
% \url{http://mosi.uni-saarland.de/people/michael}
}
\maketitle
\begin{abstract}
Many probabilistic inference problems such as stochastic filtering or the computation of rare event probabilities require model analysis under 
initial and terminal constraints. 
We propose a solution to this \emph{bridging problem}
for the widely used class of population-structured Markov jump processes.
The method is based on a
state-space lumping scheme that aggregates states in a grid structure.
The resulting approximate bridging distribution is used to iteratively refine relevant
and
truncate irrelevant parts of the state-space.
This way, the algorithm learns a well-justified finite-state
projection yielding guaranteed lower bounds for 
the system behavior under endpoint constraints.
We demonstrate the method's applicability to a wide range of problems such as Bayesian inference and the analysis of rare events.
\end{abstract}
\section{Introduction}
Discrete-valued continuous-time
 Markov Jump Processes  (MJP) are widely used to model the time evolution of complex  discrete phenomena in continuous time.
Such problems naturally occur in a wide range of areas such as chemistry~\cite{gillespie1977exact}, systems biology~\cite{wilkinson2018stochastic,ullah2011stochastic}, epidemiology~\cite{mode2000stochastic} as well as    queuing systems~\cite{breuer2003markov} and finance~\cite{pardoux2008markov}.
% The Markov property renders model analysis   feasible but requires a complete description of the current state
% to describe the future behavior of the chain.
In many applications, an MJP describes the stochastic interaction of populations of agents.
%and therefore exhibits a population structure.
The state variables are counts of individual entities of different populations.

% Observations of the MJP are typically based on noisy measurements.
% In most applications, they give only partial information about the system state and are sparse in time. 
% For example, snapshots of the spread of an epidemic are typically not accurate due to false negative/positive results and limited testing. 
 
Many tasks, such as the analysis of rare events or the inference of agent counts under partial observations naturally introduce terminal constraints on the system.
In these cases, the system's initial state is known, as well as the system's (partial) state at a later time-point.
% Parameter inference and model analysis conditioned on such sparse and partial observations are notoriously difficult as it requires endpoint conditioned simulation of the system.
% Essentially, a model must be  analyzed  with initial conditions consistent with the observations at one time-point and under the additional constraint that its subsequent evolution is consistent with observations at later times.
The probabilities corresponding to this so-called \emph{bridging problem} are often referred to as \emph{bridging probabilities} \cite{golightly2019efficient,golightly2011bayesian}.
For instance, if the exact, full state of the process $X_t$ has been observed at time $0$ and $T$, the bridging distribution is given by
$$\Pr(X_t=x\mid X_{0}=x_{0},X_{T}=x_g)$$
for all states $x$ and times $t\in [0,T]$.
Often, the condition is more complex, such that in addition to an initial distribution, a terminal distribution is present.
Such problems typically arise in a Bayesian setting, where the a priori behavior of a system is filtered such that the posterior behavior is compatible with noisy, partial observations~\cite{broemeling2017bayesian,huang2016reconstructing}.
For example, time-series data of protein levels is available while the mRNA concentration is not  \cite{adan2017flow,huang2016reconstructing}.
In such a scenario our method can be used to identify a good truncation to analyze the probabilities of mRNA levels.
% Such measurements are, for example, present in biological flow-cytometry \cite{adan2017flow} data.
% This technique allows one to measure protein concentrations at specific time points based on fluorescent labeling.
% Such measurements are noisy due to low signal intensity and can, for instance, not measure whether a gene is active or not.
% A similar problem
% \todo{finish sentence}

% In hidden Markov models  the backward-forward algorithm 
% decomposes the bridging problem into a forward and a backward
% analysis to calculate the marginal likelihood and calibrate the model according to noisy observations. \vw{haben wir da ein Zitat mit population structure und contin.-time? Hier das ist eigentlich eine HMM \cite{andreychenko2012approximate} }\todo{Luca: nicely written, but what is the link with our framework?}

Bridging probabilities also appear in the context
of rare events.
Here, the rare event is the terminal constraint because we are only interested in paths containing the event.
Typically researchers have to resort to Monte-carlo simulations in combination with variance reduction techniques in such cases~\cite{daigle2011automated,kuwahara2008efficient}.
% For instance, we may want to
% compute the probability of the rare event that the number of molecules of a certain chemical species reaches a certain cellular decision threshold \cite{daigle2011automated,kuwahara2008efficient}.

Efficient  numerical approaches  that are not based on sampling or ad-hoc approximations have rarely been developed.

Here, we combine state-of-the-art truncation strategies based on a forward analysis~\cite{lapin2011shave,andreychenko2011parameter} with a refinement approach that starts from an abstract MJP
with lumped states.
We base this lumping on a grid-like partitioning of the state-space.
Throughout a lumped state, we assume a uniform distribution that gives an efficient and convenient abstraction of the original MJP.
Note that the lumping does not follow the classical paradigm of Markov chain lumpability~\cite{buchholz1994exact}
or its variants \cite{dayar1997quasi}.
Instead of an approximate block structure of the transition-matrix used in that context, we base our partitioning on a segmentation of the molecule counts.
Moreover, during the iterative refinement of our
abstraction, we identify those regions of
the state-space that contribute most to the
bridging distribution.
In particular, we refine those lumped states that have a 
bridging probability above a certain threshold $\delta$ and
truncate all other macro-states. 
This way, the algorithm learns a truncation capturing most of the bridging probabilities. 
This truncation provides guaranteed lower bounds because it is at the granularity of the original model.

In the rest of the paper, after presenting related work (Section~\ref{sec:related}) and background (Section~\ref{sec:prelim}), we discuss the method (Section~\ref{sec:method}) and several applications, including the computation of rare event probabilities as well as Bayesian smoothing and filtering (Section~\ref{sec:results}).  

\section{Related Work}\label{sec:related}
The problem of endpoint constrained analysis occurs in the context of Bayesian estimation~\cite{sarkka2013bayesian}. 
For population-structured MJPs, this problem has been addressed by Huang et al.~\cite{huang2016reconstructing} using moment closure approximations and by Wildner and K\"{o}ppl~\cite{wildner2019moment} further employing variational inference.
Golightly and Sherlock modified stochastic simulation algorithms to approximatively  
augment generated trajectories  \cite{golightly2019efficient}.
% The idea is to  sample trajectories that are consistent with measurements, which requires sampling   appropriate values for latent state variables and sampling appropriate events between two observation times. 
Since a statistically exact augmentation is only possible for few simple cases, diffusion approximations \cite{golightly2005bayesian} and moment approximations \cite{milner2013moment} have been employed.
Such approximations, however, do not give any guarantees on the approximation error
and may suffer from numerical instabilities~\cite{schnoerr2014validity}.

The bridging problem also arises during the estimation of first passage times and rare event analysis.
Approaches for first-passage times are often of heuristic nature ~\cite{schnoerr2017efficient,hayden2012fluid,bortolussi2014stochastic}. Rigorous approaches yielding guaranteed bounds are currently limited by the performance of state-of-the-art optimization software~\cite{backenkohler2019bounding}.
In biological applications, rare events of interest are typically related to the reachability of certain thresholds on molecule counts   or mode switching \cite{strasser2012stability}.
Most methods for the estimation of rare event probabilities  rely on  importance sampling~\cite{kuwahara2008efficient,daigle2011automated}.
For other queries, alternative variance reduction techniques such as control variates are available~\cite{backenkohler2019control}. Apart from sampling-based approaches, dynamic finite-state projections have been employed by Mikeev et al.~\cite{mikeev2013numerical}, but are lacking automated truncation schemes.

The analysis of countably infinite state-spaces is often handled by a pre-defined truncation~\cite{kwiatkowska2011prism}.
Sophisticated state-space truncations for the (unconditioned) forward analysis have been developed to give lower bounds and rely on a trade-off between computational load and tightness of the bound~\cite{munsky2006finite,lapin2011shave,andreychenko2011parameter,henzinger2009sliding,mikeev2013fly}.

Reachability analysis, which is relevant in the context of probabilistic verification \cite{bortolussi2014stochastic,neupane2019stamina}, is a bridging problem where the endpoint constraint is the visit of a set of goal states.
Backward probabilities are commonly used to compute reachability likelihoods \cite{amparore2013backward,zapreev2006safe}.
Approximate techniques for reachability, based on moment closure and stochastic approximation, have also been developed in \cite{bortolussi2014stochastic,Bortolussi18infcomp}, but lack error guarantees. 
There is also a conceptual similarity between computing bridging probabilities and the forward-backward algorithm for computing state-wise posterior marginals in hidden Markov models (HMMs) \cite{rabiner1986introduction}. Like MJPs, HMMs are a generative model that can be conditioned on observations. We only consider two observations (initial and terminal state) that are not necessarily noisy but the forward and backward probabilities admit the same meaning.

\section{Preliminaries}\label{sec:prelim}
\subsection{Markov Jump Processes with Population Structure}
A population-structured Markov jump process (MJP)
describes the stochastic interactions
among agents of distinct types in a well-stirred reactor.
The assumption of all agents being equally distributed in space,
allows   to only keep track of the overall copy number of agents for each type.
Therefore the state-space is $\mathcal{S}\subseteq\mathbb{N}^{n_S}$ where
$n_S$ denotes the number of agent types or populations.
Interactions between agents are expressed as \emph{reactions}.
These reactions have associated
gains and losses of agents, given by non-negative integer vectors
${v}_j^{-}$ and ${v}_j^{+}$ for reaction $j$, respectively.
The overall effect is given by $v_j = {v}_j^{+} - {v}_j^{-}$.
A reaction between agents of types $S_1,\dots, S_{n_S}$ is specified in the following form:
\begin{equation}\label{eq:reaction}
    \sum_{\ell=1}^{n_S} v_{j\ell}^{-} S_\ell
    \xrightarrow{\alpha_j( x)}
    \sum_{\ell=1}^{n_S} v_{j\ell}^{+} S_\ell\,.
\end{equation}
The propensity function $\alpha_j$ gives the rate of the exponentially distributed firing
time of the reaction as a function of the current system state $x\in \mathcal{S}$.
In population models, \emph{mass-action} propensities are most common.
In this case the firing rate is given by the product of the number
of reactant combinations in $ x$ and a
\emph{rate constant} $c_j$, i.e.
\begin{equation}\label{eq:stoch_mass_action}
    \alpha_j({x})\coloneqq c_j\prod_{\ell=1}^{n_S}\binom{x_\ell}{v_{j\ell}^{-}}\,.
\end{equation}
In this case, we give the rate constant in \eqref{eq:reaction} instead of the   function $\alpha_j$.
For a given set of $n_R$ reactions, we define a stochastic
process $\{{{X}}_t\}_{t\geq 0}$ describing the evolution of the population
sizes over time $t$.
Due to the assumption of exponentially distributed firing times,  $ X$ is
a continuous-time
Markov chain (CTMC) on $\mathcal{S}$ with infinitesimal  generator matrix $Q$, where
the entries of $Q$ are
\begin{equation}\label{eq:cme_generator}
    Q_{ x,  y} = \begin{cases}
        \sum_{j: x+ v_j = y}\alpha_j( x)\,,&\text{if}\; x\neq
         y,\\[1ex]
        -\sum_{j=1}^{n_R} \alpha_j( x)\,, &\text{otherwise.}
    \end{cases}
\end{equation}
The probability distribution over time can be analyzed as an
initial value problem.
Given an initial state $x_0$, the distribution\footnote{In the sequel, $x_i$ denotes a state with index $i$ instead of its $i$-th component. }
\begin{equation}\label{eq:forw_prob}
\pi(x_i, t)=\Pr(X_t=x_i\mid X_0=x_0),\quad t\geq 0
\end{equation}
evolves according to the Kolmogorov forward equation
\begin{equation}\label{eq:forward}
\frac{d}{dt}\pi(t) = \pi(t) Q\,,
\end{equation}
where $\pi(t)$ is an arbitrary vectorization $(\pi(x_1,t), \pi(x_2,t),\dots,\pi(x_{|\mathcal{S}|},t))$ of the states.
% Equation~\ref{eq:forw_prob} given
% for a single state, in the context of quantitative biology, it is commonly referred to
% as the \emph{chemical master equation} (CME)\MB{can be removed if we need space}
% \begin{equation}\label{eq:cme}
%     \frac{d\pi}{d t} ( x,t) =
%     \sum_{j=1}^{n_R}\left(
%         \alpha_j( x- v_j)\pi( x- v_j,t) - \alpha_j( x)\pi( x,t)
%     \right)\,.
% \end{equation}

Let $x_g\in \mathcal{S}$ be a fixed goal state.
Given the terminal constraint $\Pr(X_T=x_g)$ for some $T\geq 0$,  we are interested in the so-called backward probabilities
\begin{equation}\label{eq:back_probs}
\beta(x_i, t) = \Pr(X_T=x_g\mid X_t = x_i),\quad t\leq T\,.
\end{equation}
Note that $\beta(\cdot, t)$ is a function of the conditional event and thus is no probability distribution over the state-space.
Instead $\beta(\cdot, t)$ gives the reaching probabilities for all states over the
time span of $[t, T]$.
To compute these probabilities, we can employ the Kolmogorov backward equation
\begin{equation}\label{eq:backward}
\frac{d}{dt}\beta(t) = Q\beta(t)^{\T}\,,
\end{equation}
where we use the same vectorization to construct $\beta(t)$ as we used
for $\pi(t)$.
The above equation is integrated backwards in time and yields the reachability
probability for each state $x_i$ and time $t<T$ of ending up in $x_g$ at time $T$.
% Similar to the CME \eqref{eq:cme}, we can state a backward chemical master equation\MB{can be removed if we need space}
% \begin{equation}\label{eq:bcme}
%     \frac{d\beta}{dt}({x}, t) =
%     \sum_{j=1}^{n_R}\left(
%         \beta( x,t) - \beta( x+ v_j,t)
%     \right)\alpha_j({x})\,.
% \end{equation}

The state-space of many MJPs with population structure, even simple ones, is countably infinite.
In this case, we have to truncate the state-space to a \emph{reasonable}
finite subset.
The choice of this truncation heavily depends on the goal of the
analysis.
If one is interested in the most ``common'' behavior, for example,
a dynamic mass-based truncation scheme is most appropriate \cite{mikeev2019approximate}.
Such a scheme truncates states with small probability during the numerical integration.
However, common mass-based truncation schemes are not as useful for the
bridging problem. This is because trajectories that meet
the specific terminal constraints can be far off the main bulk of the
probability mass.
We solve this problem by a state-space lumping in connection with an
iterative refinement scheme.

Consider as an example a birth-death process. This model can be used to model a wide variety of phenomena and often constitutes a sub-module of larger models.
For example, it can be interpreted as an M/M/1 queue with service rates being linearly dependent on the queue length.
Note, that even for this simple model, the state-space is countably infinite.
% \MB{could remove model environment for space}
\begin{model}[Birth-Death Process]\label{model:bd}
The model consists of exponentially distributed arrivals and service times proportional to queue length. It can be expressed using two mass-action reactions:
$$ \varnothing \xrightarrow{10} X \qquad\text{and}\qquad X \xrightarrow{.1} \varnothing\,.$$
The initial condition $X_0=0$ holds with probability one.
\end{model}

\subsection{Bridging Distribution}
The process' probability distribution given both initial and terminal constraints is formally described  by
the conditional probabilities
\begin{equation}\label{eq:bridge_dist}
    \gamma(x_i, t) = \Pr(X_t = x_i \mid X_0 = x_0, X_T = x_g), \quad 0\leq t\leq T
\end{equation}
for fixed initial state $x_0$ and terminal state $x_g$.
We call these probabilities the \emph{bridging probabilities}.
It is straight-forward to see that      $\gamma$ admits the
factorization
\begin{equation}\label{eq:bridge_fact}
    \gamma(x_i, t) = \pi(x_i, t)\beta(x_i, t)/\pi(x_g, T)
\end{equation}
due to the Markov property.
The normalization factor, given by the reachability probability $\pi(x_g, T)=\beta(x_0, 0)$, ensures that $\gamma(\cdot, t)$ is 
a distribution for all time points $t\in[0,T]$.
We call each $\gamma(\cdot, t)$ a   \emph{bridging distribution}.
From the Kolmogorov equations \eqref{eq:forward} and \eqref{eq:backward}
we can obtain both the forward probabilities $\pi(\cdot, t)$ and the backward probabilities
$\beta(\cdot, t)$ for $t< T$.

We can easily extend this procedure to deal with hitting times
constrained by a finite time-horizon by making the goal state $x_g$ absorbing.

\begin{figure}[tb]
    \centering
    \includegraphics[width=\textwidth]{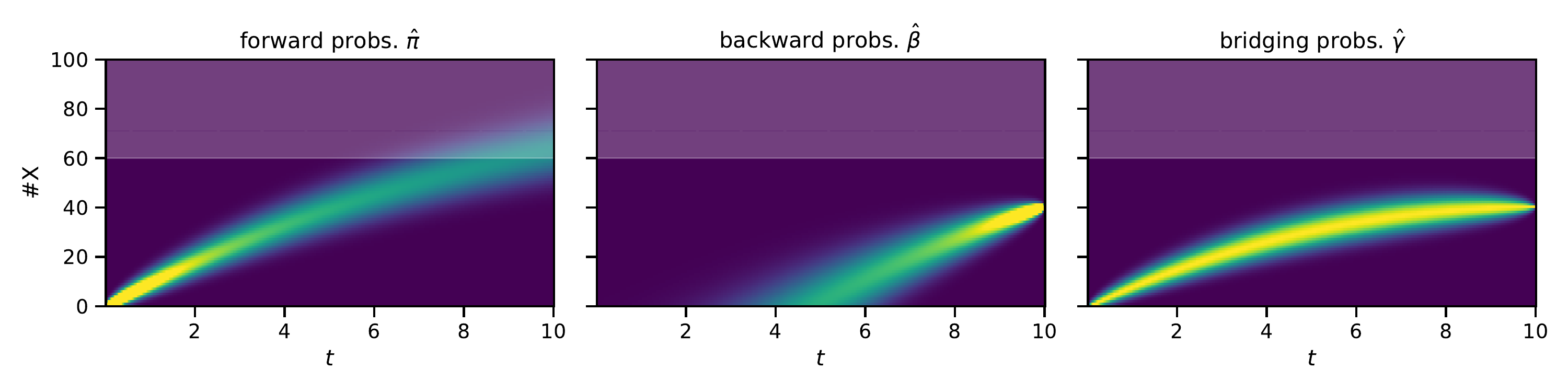}
    \caption{Forward, backward, and bridging probabilities for Model~\ref{model:bd} with initial constraint $X_0=0$ and terminal constraint $X_{10}=40$ on a truncated state-space. Probabilities over $0.1$ in $\hat\pi$ and $\hat\beta$ are given full intensity for visual clarity. %$\hat\gamma$ is normalized using the reaching probability.
    The lightly shaded area ($\geq 60$) indicates a region being more relevant for the forward than for the bridging probabilities.}
    \label{fig:bd_bridge}
\end{figure}

In Figure~\ref{fig:bd_bridge} we plot the forward, backward, and bridging probabilities for Model~\ref{model:bd}.
The probabilities are computed on a $[0,100]$ state-space truncation.
The approximate forward solution $\hat\pi$ shows how the probability mass drifts upwards towards the stationary distribution $\text{Poisson}(100)$. The backward probabilities are highest for states below the goal state $x_g=40$.
This is expected because upwards drift makes reaching $x_g$ more probable for ``lower'' states.
Finally, the approximate bridging distribution $\hat\gamma$ can be recognized to be proportional to the product of forward $\hat\pi$ and backward probabilities $\hat\beta$.

\section{Bridge Truncation via Lumping Approximations}\label{sec:method}
We first discuss the truncation of countably infinite state-spaces
to analyze backward and forward probabilities (Section~\ref{sec:fsp}).
To identify effective truncations we employ a lumping scheme.
In Section~\ref{sec:lumping}, we explain the construction of
macro-states and assumptions made, as well as the
efficient calculation of transition rates between them.
Finally, in Section~\ref{sec:alg} we present an iterative refinement algorithm
yielding a suitable truncation for the bridging problem.

\subsection{Finite State Projection}\label{sec:fsp}
Even in simple models such as a birth-death Process (Model~\ref{model:bd}), the
reachable state-space is countably infinite.
Direct analyzes of backward \eqref{eq:back_probs} and forward equations \eqref{eq:forw_prob} are often infeasible.
Instead, the integration of these differential equations requires working
with a finite subset of the infinite state-space \cite{munsky2006finite}.
If states are truncated, their incoming transitions from states that are not truncated can be re-directed to a \emph{sink state}.
The  accumulated probability in this sink state is then used
as an error estimate for the forward integration scheme.
Consequently, many truncation schemes, such as dynamic truncations \cite{andreychenko2011parameter}, aim to minimize the
amount of ``lost mass'' of the forward probability.
We use the same truncation method but base the truncation on bridging probabilities rather than the forward probabilities.
% We redirect incoming transitions of truncated states to the sink state.
% For all truncated states, we capture the
% corresponding incoming probability mass in a single sink state.

\subsection{State-Space Lumping}\label{sec:lumping}
When dealing with bridging problems, the most likely  trajectories  from the initial to the terminal state are typically not known a priori.
Especially if the event in question is rare, obtaining a state-space truncation adapted to its constraints is difficult.
We devise a lumping scheme that groups nearby states, i.e.\ molecule
counts, into larger \emph{macro-states}.
A macro-state is a collection of states treated as one state in
a lumped model, which can be seen as an abstraction of the original model.
These macro-states form a partitioning of the state-space.
In this lumped model, we assume a uniform distribution over the constituent micro-states inside each macro-state.
Thus, given that the system is in a particular macro-state, all of its micro-states are equally likely.
This partitioning allows us to analyze significant regions of the
state-space efficiently albeit under a rough approximation
of the dynamics.
Iterative refinement of the state-space after each analysis  moves the
dynamics closer to the original model.
In the final step of the iteration, 
the considered system states are at the granularity of the original model such that no approximation error is introduced by   assumptions of the lumping scheme.
Computational efficiency is retained by truncating in each iteration step those states that  contribute little probability mass to the (approximated) bridging
distributions.

We choose a lumping scheme based on a grid of hypercube macro-states whose endpoints belong to a predefined grid.
This topology makes the computation of transition rates between macro-states
particularly convenient.
Mass-action reaction rates, for example, can be given in a closed-form
due to the Faulhaber formulae.
More complicated rate functions such as Hill functions can often be handled as well by taking appropriate integrals.

Our choice is a scheme that uses $n_S$-dimensional hypercubes.
% A macro-state can therefore be described as the cross-product of
% intervals over $\mathbb{N}$:
% \begin{equation}
% \bar{x}_i = \left([\ell_i^{(1)}, u_i^{(1)}]\times
% \dots\times [\ell_i^{(n_S)}, u_i^{(n_S)}]
% \right)\bigcap \mathbb{N}^{n_S}
% \end{equation}
A macro-state $\bar{x}_i(\ell^{(i)},u^{(i)})$ (denoted by $\bar{x}_i$ for notational ease) can therefore be described by two vectors $\ell^{(i)}$
and $u^{(i)}$.
The vector $\ell^{(i)}$ gives the corner closest to the origin, while $u^{(i)}$
gives the corner farthest from the origin.
Formally,
\begin{equation}\label{eq:macros_state}
    \bar{x}_i = \bar{x}_i(\ell^{(i)},u^{(i)}) =  \{x\in\mathbb{N}^{n_S} \mid  \ell^{(i)}  \leq x  \leq u^{(i)} \},
\end{equation}
where '$\leq$' stands for the element-wise comparison.
This choice of topology makes the computation of transition
rates between macro-states particularly convenient:
Suppose we are interested in the set of micro-states
in macro-state $\bar{x}_i$ that can transition
to macro-state $\bar{x}_k$ via reaction $j$.
It is easy to see that this set is itself an
interval-defined macro-state $\bar{x}_{i\xrightarrow{j}k}$.
To compute this macro-state
we can simply shift $\bar{x}_i$ by $v_j$, take the intersection
with $\bar{x}_k$ and project this set back.
Formally,
\begin{equation}\label{eq:transition_set}
    \bar{x}_{i\xrightarrow{j}k} = ((\bar{x}_i + v_j) \cap \bar{x}_k) - v_j\,,
\end{equation}
where the additions are applied element-wise to all states
making up the macro-states.
For the correct handling of the truncation it is useful to
define a general exit state
\begin{equation}
    \bar{x}_{i\xrightarrow{j}} = ((\bar{x}_i + v_j) \setminus \bar{x}_i) - v_j.
\end{equation}
This state captures all micro-states inside $\bar{x}_i$ that can leave the state via reaction $j$.
Note that all operations preserve the structure of a macro-state as defined in \eqref{eq:macros_state}.
Since a macro-state is based on intervals the computation
of the transition rate is often straight-forward.
Under the assumption of polynomial rates, as it is the case for mass-action
systems, we can compute the sum of rates over this transition set
efficiently using Faulhaber's formula.
We define  the lumped transition function
\begin{equation}\label{eq:lumped_propfun}
    \bar{\alpha}_j({\bar{x}}) = \sum_{x\in \bar{x}} \alpha_j(x)
\end{equation}
for macro-state $\bar{x}$ and reaction $j$.
As an example consider the following mass-action reaction
$ 2 X \xrightarrow{c} \varnothing\,. $
For macro-state
$\bar{x} = \{0, \dots, n\}$
we can compute the corresponding lumped transition rate
$$\bar{\alpha}(\bar{x})
=\frac{c}{2}\sum_{i=1}^n i (i - 1)
=\frac{c}{2}\sum_{i=1}^n (i^2 - i)
=\frac{c}{2}\left(\frac{2n^3+3n^2+n}{6} - \frac{n^2 + n}{2}\right)$$
eliminating the explicit summation in the lumped propensity function.

For polynomial propensity functions $\alpha$ such formulae
are easily obtained automatically.
For non-polynomial propensity functions, we can use
the continuous integral as an approximation.
This is demonstrated on a case study in Section~\ref{sec:hill_toggle}.
\begin{figure}[t]
    \begin{minipage}{0.49\textwidth}
    \centering
    \includegraphics[scale=.39]{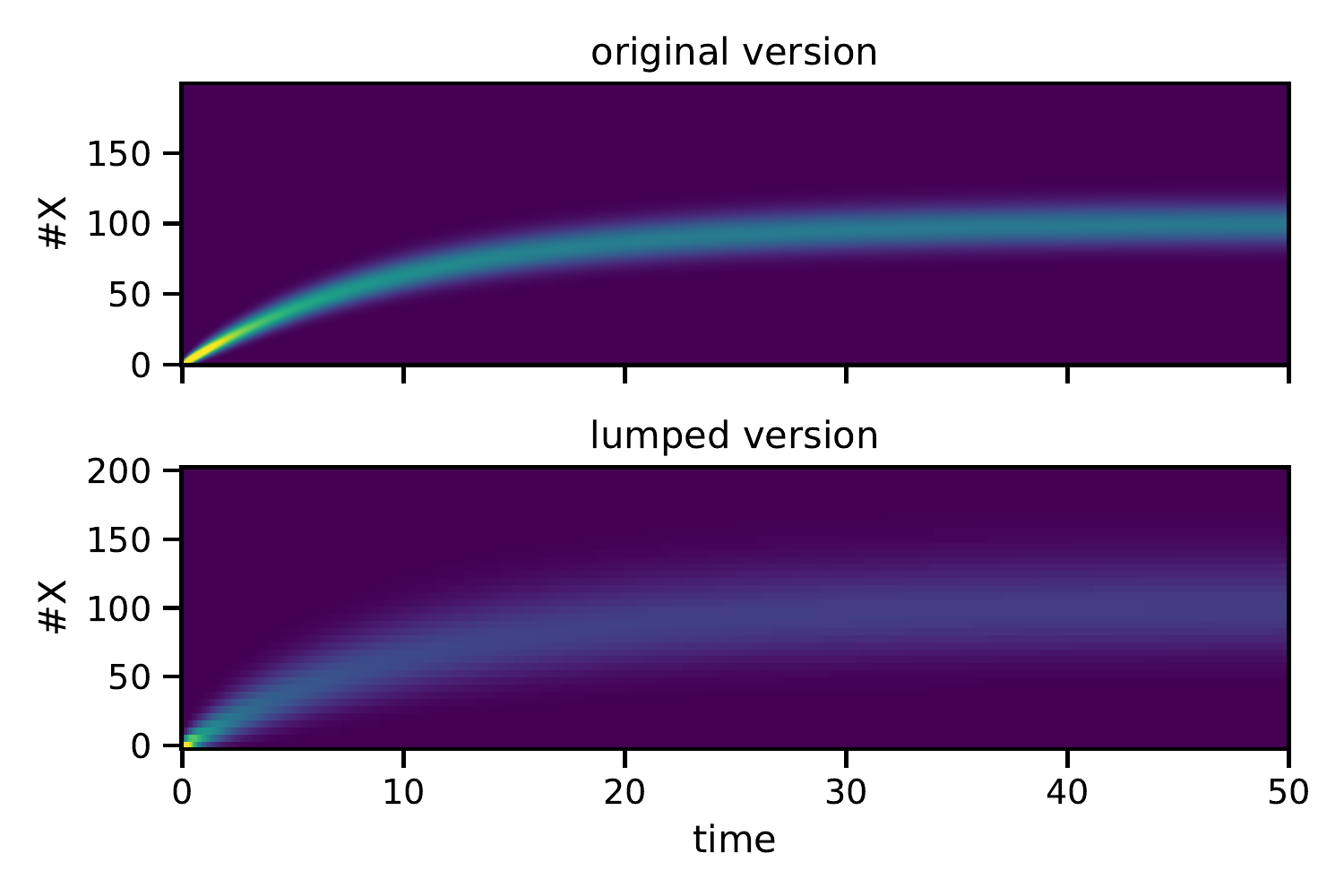}
    \end{minipage}
    \hfill
    \begin{minipage}{0.49\textwidth}
    \centering
    \includegraphics[scale=.39]{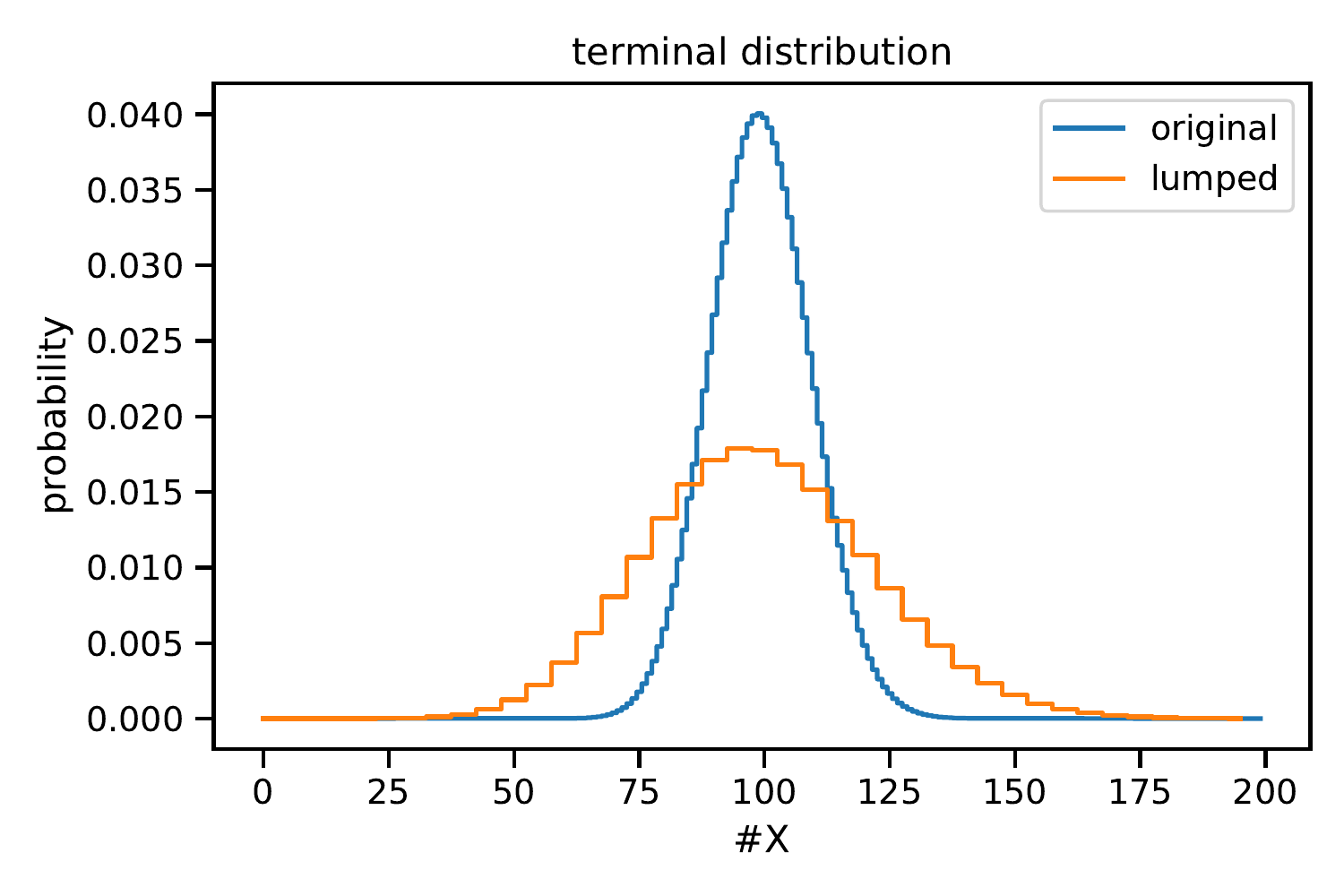}
    \end{minipage}
    \caption{A lumping approximation of Model~\ref{model:bd} on the state-space truncation to $[0, 200]$ on $t\in[0, 50]$. On the left-hand side solutions of a regular truncation approximation and a lumped truncation (macro-state size is 5) are given. On the right-hand side the respective terminal distributions $\Pr(X_{50}=x_i)$ are contrasted.}
    \label{fig:lumped}
\end{figure}

Using the transition set computation \eqref{eq:transition_set}
and the lumped propensity function \eqref{eq:lumped_propfun}
we can populate the $Q$-matrix of the finite lumping approximation:
\begin{equation}\label{eq:lumped_q}
    \bar{Q}_{ \bar{x}_i,  \bar{x}_k} = \begin{cases}
        \sum_{j=1}^{n_R}{\bar\alpha}_j\left(\bar{x}_{i\xrightarrow{j}k}\right)/{\mathrm{vol}\left(\bar{x}_i\right)}\,,&\text{if}\; \bar{x}_i\neq \bar{x}_k\\[1ex]
        -\sum_{j=1}^{n_R}{\bar\alpha}_j\left(\bar{x}_{i\xrightarrow{j}}\right)/{\mathrm{vol}\left(\bar{x}_i\right)}\,, &\text{otherwise}
    \end{cases}
\end{equation}
In addition to the lumped rate function over the transition state $\bar{x}_{i\xrightarrow{j}k}$, we need to divide by the total volume of the lumped state $\bar{x}_i$.
This is due to the assumption of a uniform distribution inside the macro-states.
Using this $Q$-matrix, we can compute the forward and backward solution
using the respective Kolmogorov equations \eqref{eq:forward} and
\eqref{eq:backward}.

Interestingly, the lumped distribution
tends to be less concentrated. %\MB{rephrase this passage in terms of entropy?}\todo{L: this is interesting, but why? maybe an intuition of why uniform spreads more the distribution}
This is due to the assumption of a
uniform distribution inside macro-states.
This effect is illustrated by the example of a birth-death process in Figure~\ref{fig:lumped}.
Due to this effect, an iterative refinement typically keeps an over-approximation in terms of state-space area.
This is a desirable feature since relevant regions are less likely to be pruned due to lumping approximations.

\subsection{Iterative Refinement Algorithm}\label{sec:alg}
The iterative refinement algorithm (Alg.~\ref{alg:refinement}) starts with a set of large macro-states
that are iteratively refined, based on approximate solutions
to the bridging problem.
We start by constructing square macro-states of size
$2^m$ in each dimension for some $m\in\mathbb{N}$ such that they form a large-scale grid $\mathcal{S}^{(0)}$.
Hence, each initial macro-state has a volume of ${\left(2^m\right)}^{n_S}$.
This choice of grid size is convenient because we can halve states
in each dimension.
Moreover, this choice ensures that all states have equal volume
and we end up with states of volume $2^0=1$ which is
equivalent to a truncation of the original non-lumped state-space.\\
\indent An iteration of the state-space refinement starts by computing both the
forward and backward probabilities (lines \ref{line:forw} and \ref{line:backw})  via integration of \eqref{eq:forward} and
\eqref{eq:backward}, respectively, using
 the lumped $\hat Q$-matrix.
Based on the resulting approximate forward and backward probabilities, we
  compute an approximation of the
bridging distributions (line~\ref{line:bridge}).
This is done for each time-point in an equispaced grid on $[0,T]$.
The time grid granularity is a hyper-parameter of the algorithm.
If the grid is too fine, the memory overhead of storing backward $\hat\beta^{(i)}$
and forward solutions $\hat\pi^{(i)}$ increases.\footnote{We denote the approximations with a hat (e.g.\ $\hat{\pi}$) rather than a bar (e.g.\ $\bar{\pi}$) to indicate that not only the lumping approximation but also a truncation is applied and similarly for the $Q$-matrix.}
If, on the other hand, the granularity is too low, too much of
the state-space might be truncated.
Based on a threshold parameter $\delta>0$
states are either removed or split (line~\ref{line:refine}), depending on
the mass assigned to them by the approximate bridging
probabilities $\hat\gamma^{(i)}_t$.
A state can be split by the \texttt{split}-function which
halves the state in each dimension.
Otherwise, it is removed.
Thus, each macro-state is either split into $2^{n_S}$ new states or removed
entirely.
The result forms the next lumped state-space $\mathcal{S}^{(i+1)}$.
 The   $Q$-matrix is adjusted (line~\ref{line:update_q}) such that transition rates  for $\mathcal{S}^{(i+1)}$  are calculated according to 
 \eqref{eq:lumped_q}. 
 Entries of truncated states are removed from the transition matrix. Transitions leading to them are
re-directed to a sink state (see Section \ref{sec:fsp}).
After $m$ iterations (we started with states of side lengths $2^m$)
we have a standard finite state projection scheme
on the original model tailored to
computing an approximation of the bridging distribution.
\begin{algorithm}[htb]
\SetKwFunction{Split}{split}
\SetKwInOut{Input}{input}
\SetKwInOut{Output}{output}
\Input{Initial partitioning $\mathcal{S}^{(0)}$, truncation threshold $\delta$}
\Output{approximate bridging distribution $\hat\gamma$}
\For{$i=1,\dots,m$}{
    ${\hat\pi}^{(i)}_t\leftarrow $ solve approximate forward equation on $\mathcal{S}^{(i)}$\label{line:forw}\;
    ${\hat\beta}^{(i)}_t\leftarrow $ solve approximate backward equation on $\mathcal{S}^{(i)}$\label{line:backw}\;
    ${\hat\gamma}^{(i)}_t\leftarrow {\hat\beta}^{(i)} {\hat\pi}^{(i)}/\hat\pi(x_g,  T)$\label{line:bridge}\tcc*{approximate bridging distribution}
    $\mathcal{S}^{(i+1)}\leftarrow \emptyset$\;
    \ForEach{$\bar{x}\in\mathcal{S}^{(i)}$}{
        \If{
            $\exists t.{\hat\gamma}^{(i)}_t(\bar{x})\geq \delta$\label{line:refine}\tcc*{refine based on bridging probabilities}
        }{
            $\mathcal{S}^{(i+1)}\leftarrow \mathcal{S}^{(i+1)} \cup \Split(\bar{x})$\label{line:union}\;
        }}
        update $\hat{Q}$-matrix\label{line:update_q}\;
}
\Return ${\hat\gamma}^{(i)}$\;
    \caption{Iterative refinement for the bridging problem}
    \label{alg:refinement}
\end{algorithm}

\begin{figure}[tb]
    \centering
    \includegraphics[width=\textwidth]{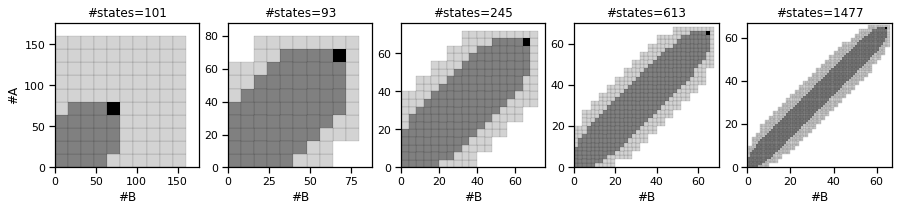}
    \caption{The state-space refinement algorithm on two parallel unit-rate arrival processes. The bridging problem from $(0,0)$ to $(64, 64)$ and $T=10$ and truncation threshold $\delta=\e{5}{-3}$. States with a bridging probability below $\delta$ are light grey. The macro-state containing the goal state is marked in black. The initial macro-states are of size $16\times 16$.}
    \label{fig:refinement}
\end{figure}

In Figure~\ref{fig:refinement} we give a demonstration of how
Algorithm~\ref{alg:refinement} works to refine the state-space
iteratively. Starting with an initial lumped state-space $\mathcal{S}^{(0)}$ covering a large area of the state-space,
repeated evaluations of the bridging distributions are performed.
After five iterations the remaining truncation includes all states
that significantly contribute to the bridging
probabilities over the times $[0,T]$.

% das ist nicht spezifisch für rare events. es kann genauso gut sein dass bei einem rare event die hauptmasse der forward in der truncation bleibt.
It is important to realize that determining the most relevant states is \emph{the} main challenge.
The above algorithm solves this problem by considering only those parts of the state-space that contribute most to the bridging probabilities.
The truncation is tailored to this condition and might ignore regions that are likely in the unconditioned case.
For instance, in Fig.~\ref{fig:bd_bridge} the bridging probabilities mostly remain below a population threshold of $\#X=60$ (as indicated by the lighter/darker coloring), while the forward probabilities mostly exceed this bound. Hence, in this example a significant portion of the forward probabilities $\hat\pi_t^{(i)}$  
 is captured by the sink state. However, the condition in line~\ref{line:refine} in Algorithm~\ref{alg:refinement} ensures that
states contributing significantly to $\hat\gamma_t^{(i)}$ will be kept and refined in the next iteration.

\section{Results}\label{sec:results}
% Three examples are chemical reaction networks whose
% stochastic evolution is described through an MJP.
% The corresponding population structure is given by the 
% copy numbers of different chemical species. 
% In this application field, an important challenge is the computation of rare event probabilities \cite{daigle2011automated,chong2017path}. 
% Rare events of interest are typically reachability of certain thresholds on molecule counts   or mode switching \cite{strasser2012stability}. <- ist jetzt im rel work
% 
We present four examples in this section to evaluate our proposed method. 
A prototype was implemented in Python~3.8. For numerical integration we used the Scipy implementation \cite{2020SciPy-NMeth} of the implicit method based on backward-differentiation formulas \cite{byrne1975polyalgorithm}.
The analysis as a Jupyter notebook is made available online\footnote{\url{https://www.github.com/mbackenkoehler/mjp_bridging}}.

\subsection{Bounding Rare Event Probabilities}

We consider a simple model of two parallel Poisson processes describing the production of two 
types of agents. 
The corresponding probability distribution
has Poisson product form at all time points $t\geq 0$ and hence we can compare the accuracy of our  numerical results with the exact analytic solution.
We use the proposed approach to compute lower bounds for rare event probabilities.
\footnote{These bounds are rigorous up to the approximation error of the numerical integration scheme. However, the forward solution could be replaced by an adaptive uniformization approach~\cite{andreychenko2010fly} for a more rigorous integration error control.}

\begin{figure}[!t]
    \centering
    \includegraphics[width=\textwidth]{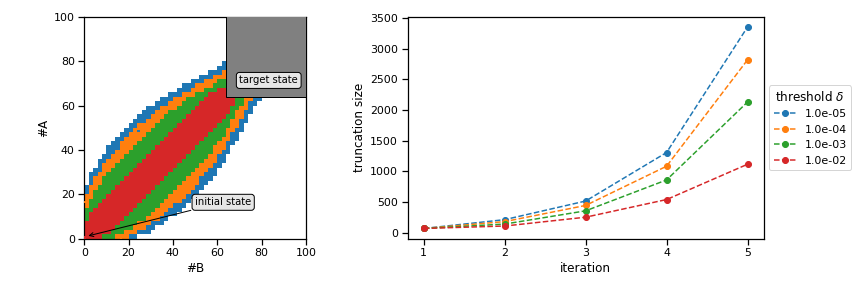}
    \caption{State-space truncation for varying values of the
    threshold parameter $\delta$: Two parallel Poisson processes under terminal constraints $X_{20}^{(A)} \geq 64$ and $X_{20}^{(B)}\geq 64$.
    The initial macro-states are $16\times 16$ such that the final
    states are regular micro states.}
    \label{fig:2poisson_rare}
\end{figure}

\begin{model}[Parallel Poisson Processes]
The model consists of two parallel independent Poisson processes with unit rates.
$$ \varnothing \xrightarrow{1} A \qquad\text{and}\qquad \varnothing \xrightarrow{1} B\,. $$
The initial condition $X_0=(0,0)$ holds with probability one. After $t$ time units each species abundance
is Poisson distributed with rate $\lambda=t$.
\end{model}
We consider the final constraint of reaching a state where both
processes exceed a threshold of 64 at time 20.
Without prior knowledge, a reasonable truncation would have been $160\times 160$. But our analysis shows that just 20\% of the states are necessary to capture over 99.6\% of the probability mass reaching the
target event (cf.\ Table~\ref{tab:par_poisson}).
Decreasing the threshold $\delta$ leads to a larger set of states retained after truncation as more of the bridging distribution is included (cf.\ Figure~\ref{fig:2poisson_rare}).
We observe an increase in truncation size that is approximately
logarithmic in $\delta$, which, in this example, indicates robustness of the method with respect to the choice of $\delta$.

\setlength{\tabcolsep}{10pt}
\begin{table}[!t]
    \centering
    \begin{tabular}{lrrrr}
    \toprule
    threshold $\delta$ & 1e-2 & 1e-3 & 1e-4 & 1e-5 \\
    \midrule
    truncation size & 1154 & 2354 & 3170 & 3898 \\
    overall states & 2074 & 3546 & 4586 & 5450 \\
    estimate& 8.8851e-30 & 1.8557e-29 & 1.8625e-29 & 1.8625e-29\\
    rel.\ error& 5.2297e-01 & 3.6667e-03 & 3.7423e-05 & 9.5259e-08\\
    \bottomrule
    \end{tabular}
    \caption{Estimated reachability probabilities based on varying truncation thresholds $\delta$: The true probability is
    1.8625e-29. We also report the size of the final truncation and the accumulated size of all truncations during refinement iterations (overall states).}
    \label{tab:par_poisson}
\end{table}

\paragraph{Comparison to other methods}
The truncation approach that we apply  is similar  to the one used by Mikeev et al.~\cite{mikeev2013numerical} for rare event estimation.
However,  they used a given linearly biased MJP model to obtain a truncation. A general strategy to compute an appropriate biasing was not proposed.
It is possible to adapt our truncation approach to the dynamic scheme in Ref.~\cite{mikeev2013numerical} where states are removed in an on-the-fly fashion during numerical integration.
% For this dynamic adaption one can change the union over all time points 
% (Algorithm~\ref{alg:refinement} line~\ref{line:union}) to a union
% over time intervals. \todo{L: I do not get this really? line 7 is the split. } % MB: this point is not important

A finite state-space truncation covering the same area as the initial lumping approximation would contain $25,\!600$ states.\footnote{Here, the goal  is not treated as a single state. Otherwise, it consists of $24,\!130$ states.}
The standard approach would be to build up the entire state-space for such a model \cite{kwiatkowska2011prism}.
Even using a conservative truncation threshold $\delta=\text{1e-5}$, our method yields an accurate estimate using only about a fifth (5450) of this accumulated over all intermediate lumped approximations. 
% Moreover the analysis yields intermediate distributions which we would not obtain using PRISM.

\subsection{Mode Switching}
%Next, we turn to ``mode switching'' events.
%These occur in models exhibiting a \emph{multi-modal} behavior, which are of great relevance in
%many contexts~\cite{siegal2011emergence}. These models exhibit a multi-modal stationary distribution.
Mode switching occurs in models exhibiting   \emph{multi-modal} behavior~\cite{siegal2011emergence} when a trajectory   traverses a potential barrier from one mode
to another.
%As such this can, in most cases, be considered a sub-class
%of rare-event analysis.
Often, mode switching is a rare event and occurs in the context of gene regulatory networks where a mode is characterized by the set of genes being currently active~\cite{loinger2007stochastic}. 
Similar dynamics also commonly occur in queuing models where a system may for example switch its operating  behavior stochastically if
traffic increases above or decreases below certain thresholds.
Using the presented method, we can get both a   qualitative and quantitative understanding of   switching
behavior without resorting to Monte-Carlo methods such as   importance sampling.
% In the most common importance sampling scheme, for example,
% rate functions are biased using linear constants, fitted according to a cross-entropy objective~\cite{daigle2011automated}.
% These biased paths are not necessarily representative of the bridging
% distributions.
%Thus we can gain a deeper insight into
%the actual mechanics of mode switching.
% \begin{itemize}
%     \item Mode Switching as a subclass of rare event problems
%     \item Switching dynamics are of biological interest (cell cycle control, stuff like that)
%     \item Very common ``sub-module'' of more complex models
%     \item Especially hard to tackle for standard FSP methods (multi-modal stationary distribution)
% \end{itemize}

\subsubsection{Exclusive Switch}
The exclusive switch~\cite{barzel2008calculation}  has three different modes of operation, depending on the
DNA state, i.e.\ on whether a protein of type one or two is bound to
the DNA.

\begin{model}[Exclusive Switch] The exclusive switch model consists of a promoter region
that can express both proteins $P_1$ and $P_2$. Both can bind to the region, suppressing
the expression of the other protein. For certain parameterizations, this leads to a
bi-modal or even tri-modal behavior.
$$ D \xrightarrow{\rho} D + P_1 \qquad D \xrightarrow{\rho} D + P_2 \qquad P_1 \xrightarrow{\lambda}\varnothing \qquad P_2 \xrightarrow{\lambda} \varnothing $$
$$ D + P_1 \xrightarrow{\beta} D.P_1 \qquad D.P1 \xrightarrow\gamma D + P_1 \qquad D.P_1 \xrightarrow\alpha D.P_1 + P_1 $$
$$ D + P_2 \xrightarrow{\beta} D.P_2 \qquad D.P2 \xrightarrow{\gamma} D + P_2 \qquad D.P_2 \xrightarrow\alpha D.P_2 + P_2 $$
The parameter values are $\rho=\e{1}{-1}$, $\lambda=\e{1}{-3}$, $\beta=\e{1}{-2}$, $\gamma=\e{8}{-3}$, and $\alpha=\e{1}{-1}$.
\end{model}
Since we know a priori of the three distinct operating modes, we adjust
the method slightly:
The state-space for the DNA states is not lumped. Instead we ``stack''
lumped approximations of the $P_1$-$P_2$ phase space upon each other. Special treatment of DNA states is common for such models \cite{lapin2011shave}. 

%We assess the switching dynamics from one mode to another.
To analyze the switching, we choose the transition from (variable order: $P_1$, $P_2$, $D$, $D.P_1$, $D.P_2$)
$ x_{1} = (32, 0, 0, 0, 1) $
to
$ x_2 = (0, 32, 0, 1, 0) $
over the time interval $t\in[0,10]$.
The initial lumping scheme covers up to 80 molecules of $P_1$ and $P_2$ for each mode.
Macro-states have size $8\times8$ and
the truncation threshold is $\delta=\e{1}{-4}$.

\begin{figure}[t]
    \begin{minipage}{.54\textwidth}
    \centering
    \includegraphics[scale=.45]{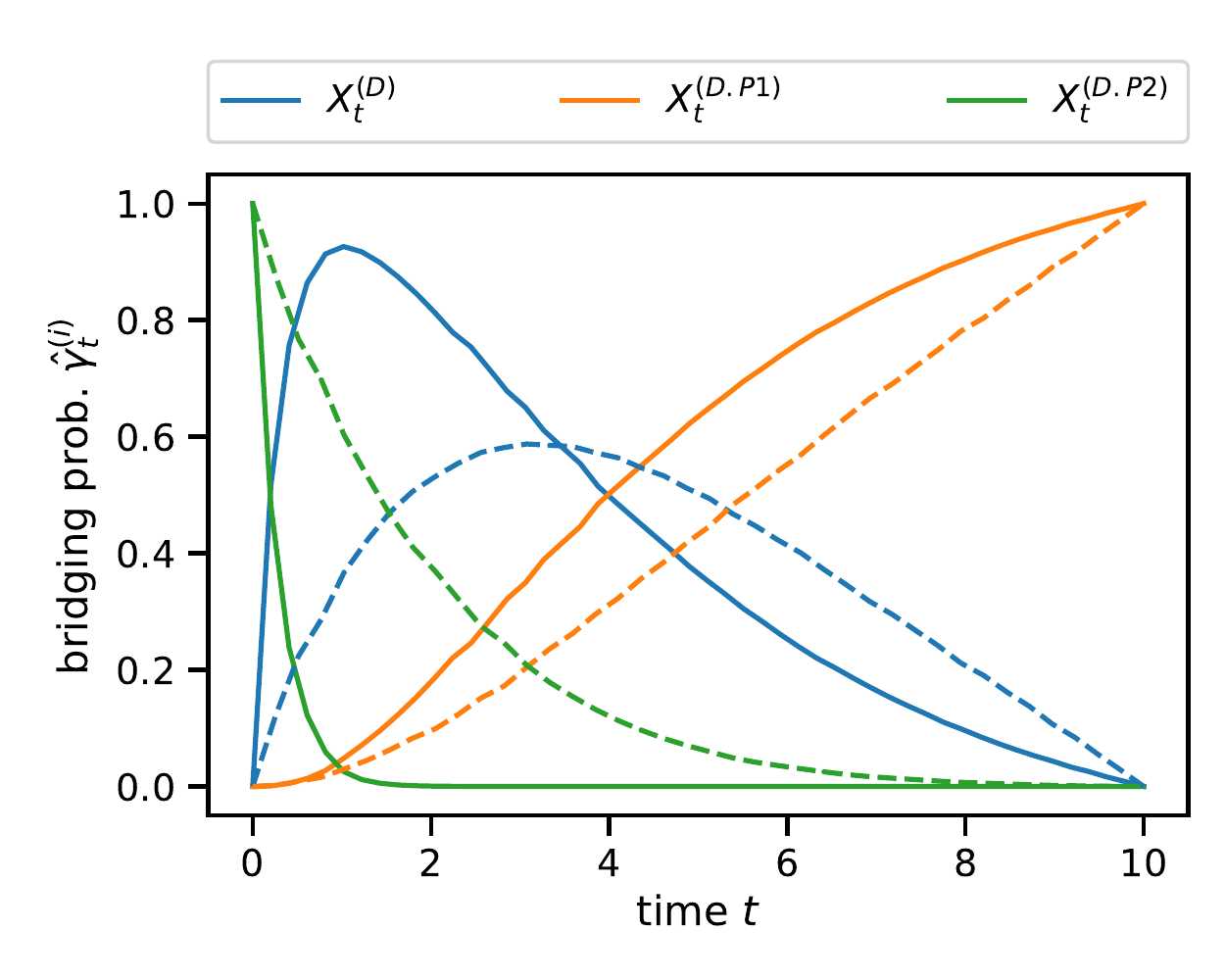}
    \end{minipage}
    \hspace{-1ex}
    \begin{minipage}{.455\textwidth}
    \centering
    \includegraphics[scale=.43]{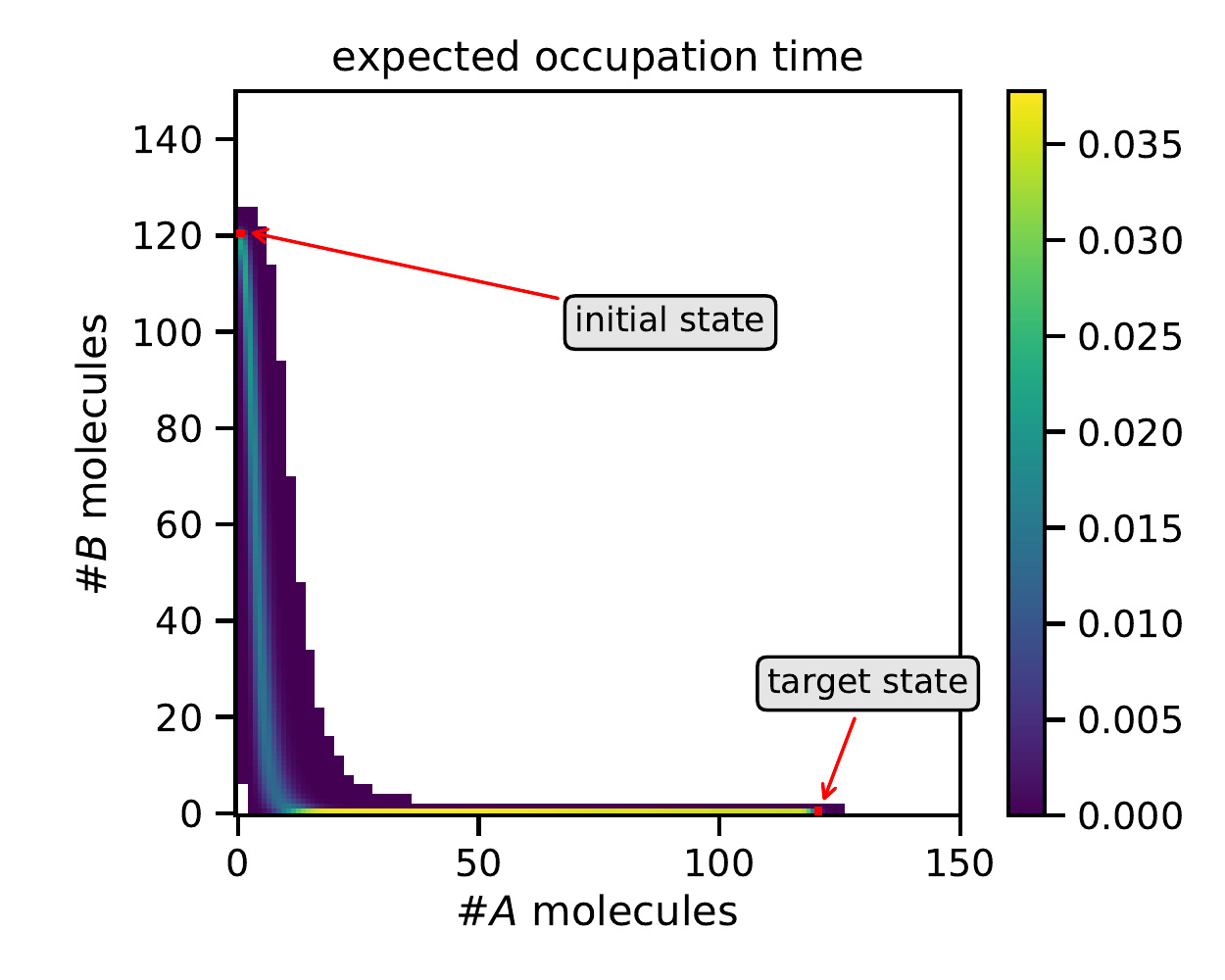}
    \end{minipage}
    \caption{(left) Mode probabilities of the exclusive switch bridging problem over time for the first lumped approximation (dashed lines) and the final approximation (solid lines) with constraints $X_0=(32, 0, 0, 1, 0)$ and $X_{10}=(0,32,0,0,1)$.  
    (right) The expected occupation time (excluding initial and terminal states) for the switching problem of the toggle switch using Hill-type functions. The bridging problem is from initial $(0,120)$ to a first passage of $(120, 0)$ in
    $t\in [0,10]$.}
    \label{fig:switching_dynamics}
\end{figure}

In the analysis of biological switches, not only the switching probability    
%is of interest to the researcher.
%The 
but also the switching dynamics is a central part of understanding the underlying biological mechanisms.
In Figure~\ref{fig:switching_dynamics} (left), we therefore plot the time-varying probabilities of the gene state conditioned on the mode.
We observe a rapid unbinding of $P_2$, followed by a slow increase of the binding probability for $P_1$.
These dynamics are already qualitatively captured by the first lumped approximation (dashed lines).
% \begin{itemize}
%     \item emphasize that we not only get rare event prob.\ estimates but insights on the dynamics leading there
%     \item explain how mode switching dynamics
%     \item Compare/contrast coarse- and fine-grained
% \end{itemize}

\subsubsection{Toggle Switch}\label{sec:hill_toggle}
Next, we apply our method to  a toggle switch model exhibiting non-polynomial rate functions.
This well-known  model considers two proteins $A$ and $B$
inhibiting the production of the respective other protein~\cite{lipshtat2006genetic}.
\begin{model}{Toggle Switch (Hill functions)}
We have population types $A$ and $B$ with the following reactions and reaction rates.
$$ \varnothing \xrightarrow{\alpha_1(\cdot)} A\,,\quad \text{where}\quad \alpha_1(x) = \frac{\rho}{1 + x_B},
\qquad A \xrightarrow\lambda \varnothing $$
$$ \varnothing \xrightarrow{\alpha_1(\cdot)} B\,,\quad \text{where}\quad \alpha_1(x) = \frac{\rho}{1 + x_A},
\qquad B \xrightarrow\lambda \varnothing $$
The parameterization is $\rho=10$, $\lambda=0.1$.
\end{model}
Due to the non-polynomial rate functions $\alpha_1$ and $\alpha_2$, 
the transition rates between macro-states
 are approximated by using the continuous integral
$$
\bar{\alpha}_1(\bar{x})\approx\int_{a-0.5}^{b+0.5} \frac{\rho}{1 + x}\, dx = \rho \left(\log{\left(b + 1.5 \right)} - \log{\left(a + 0.5 \right)} \right)
$$
for a macro-state $\bar{x}= \{a,\dots,b\}$.%\vw{explain $k$ I guess it is the mode}%\MB{we don't have mode vars here. it's just $k$ to make the expression more clear because then its essentially 1D}

We analyze the switching scenario from $(0, 120)$ to the first visit of state $(120, 0)$ up to time $T=10$. The initial lumping scheme covers up to 352 molecules of $A$ and $B$ and  macro-states have size $32\times32$.
The truncation threshold is $\delta=\e{1}{-4}$.
%Transition rates are computed using the above non-polynomial integral.
The resulting truncation is shown in Figure~\ref{fig:switching_dynamics} (right).
It also illustrates the kind of insights that can be obtained from the bridging distributions.
For an overview of the switching dynamics, we look at the expected occupation time under the terminal constraint of having entered state $(120,0)$. Letting the corresponding hitting time be $\tau=\inf\{t\geq 0\mid X_t=(120, 0)\}$, the expected occupation time for some state $x$ is
$E\left(\int_0^{\tau}1_{=x}(X_t)\,dt\mid \tau\leq 10\right)$.
We observe that in this example the switching  behavior seems to be asymmetrical.
The main mass seems to pass through an area where initially a small number of $A$ molecules is produced followed by a total decay of $B$ molecules.

\subsection{Recursive Bayesian Estimation}
We now turn to the method's application in recursive Bayesian estimation.
This is the problem of estimating the system's
past, present, and future behavior under given observations.
Thus, the MJP becomes a hidden Markov model (HMM).
The observations in such models are usually noisy, meaning that we cannot infer the system state with certainty.

This estimation problem entails more general distributional constraints on terminal $\beta(\cdot,T)$ and initial $\pi(\cdot, 0)$ distributions than the point mass distributions considered up until now.
We can easily extend the forward and backward probabilities to more general initial
distributions   and terminal distributions $\beta(T)$.
%This extension is useful for estimating scenarios subject to measurement noise.
For the forward probabilities we get
\begin{equation}
    \pi(x_i, t) = \sum_j \Pr(X_t=x_i\mid X_0=x_j) \pi(x_j,0),
\end{equation}
and similarly the backward probabilities are given by
\begin{equation}
    \beta(x_i, t) = \sum_j\Pr(X_T=x_j\mid X_t = x_i) \beta_T(x_j)\,.
\end{equation}
We apply our method to an SEIR (susceptible-exposed-infected-removed) model.
This is widely used to describe the spreading of an epidemic such as the current COVID-19 outbreak~\cite{he2020seir,grossmann2020importance}.
Temporal snapshots of the epidemic spread  are mostly only available for a subset of the population and suffer from   inaccuracies of diagnostic tests.
Bayesian estimation can then be used to infer the spreading dynamics given uncertain temporal snapshots.
%Therefore due to the combination of model simulation and the uncertain measurements a  problem arises.
%We demonstrate our method on such an estimation problem.
% We assume the full model including parameterization is given.

\begin{model}[Epidemics Model]\label{model:seir}
A population of susceptible individuals can contract a disease from infected agents. In this case, they are exposed, meaning they will become infected but cannot yet infect others. After being infected, individuals change to the removed state. The mass-action reactions are as follows.
$$ S + I \xrightarrow{\lambda} E + I \qquad
E \xrightarrow{\mu} I \qquad
I \xrightarrow{\rho} R $$
The parameter values are $\lambda=0.5$, $\mu=3$, $\rho=3$. Due to the stoichiometric invariant $X_t^{(S)} + X_t^{(E)} + X_t^{(I)} + X_t^{(R)} = \mathrm{const.}$, we can eliminate $R$ from the system.
\end{model}

We consider the following scenario:
We know that initially ($t=0$) one individual is infected and the rest is susceptible.
At time $t=0.3$ all individuals are tested for the disease.
The test, however, only identifies infected individuals with probability $0.99$.
Moreover, the probability of a false positive is 0.05.
We like to identify the distribution given both the initial state and the measurement at time $t=0.3$.
In particular, we want to infer the distribution over the latent counts of  $S$ and $E$ by
% The question we like to answer is whether the number of infected exceeds the threshold $k=100$ in the time interval $[t,2t]$.
  \emph{recursive Bayesian estimation}.

\begin{figure}[t]
    \centering
    \begin{minipage}{0.37\textwidth}
    \centering
    \includegraphics[scale=.4]{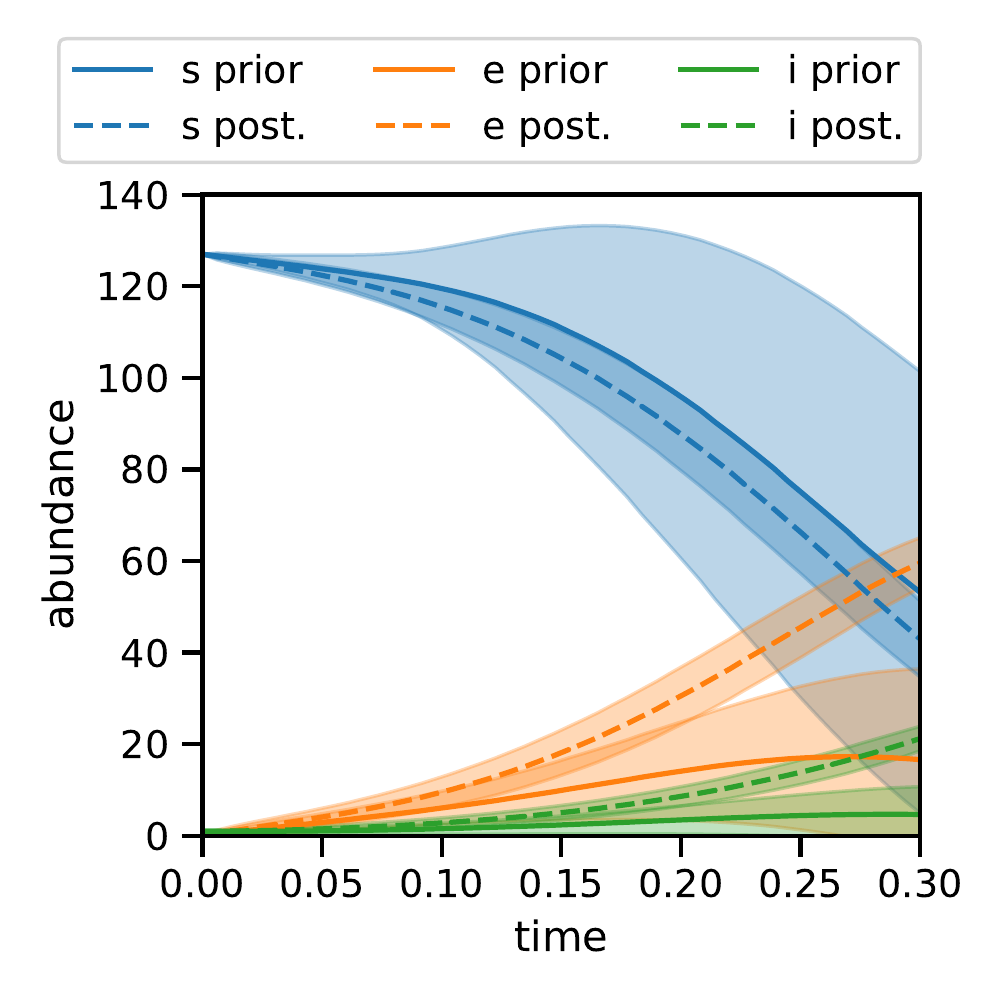}
    \end{minipage}
    \begin{minipage}{0.37\textwidth}
    \centering
    \includegraphics[scale=.4]{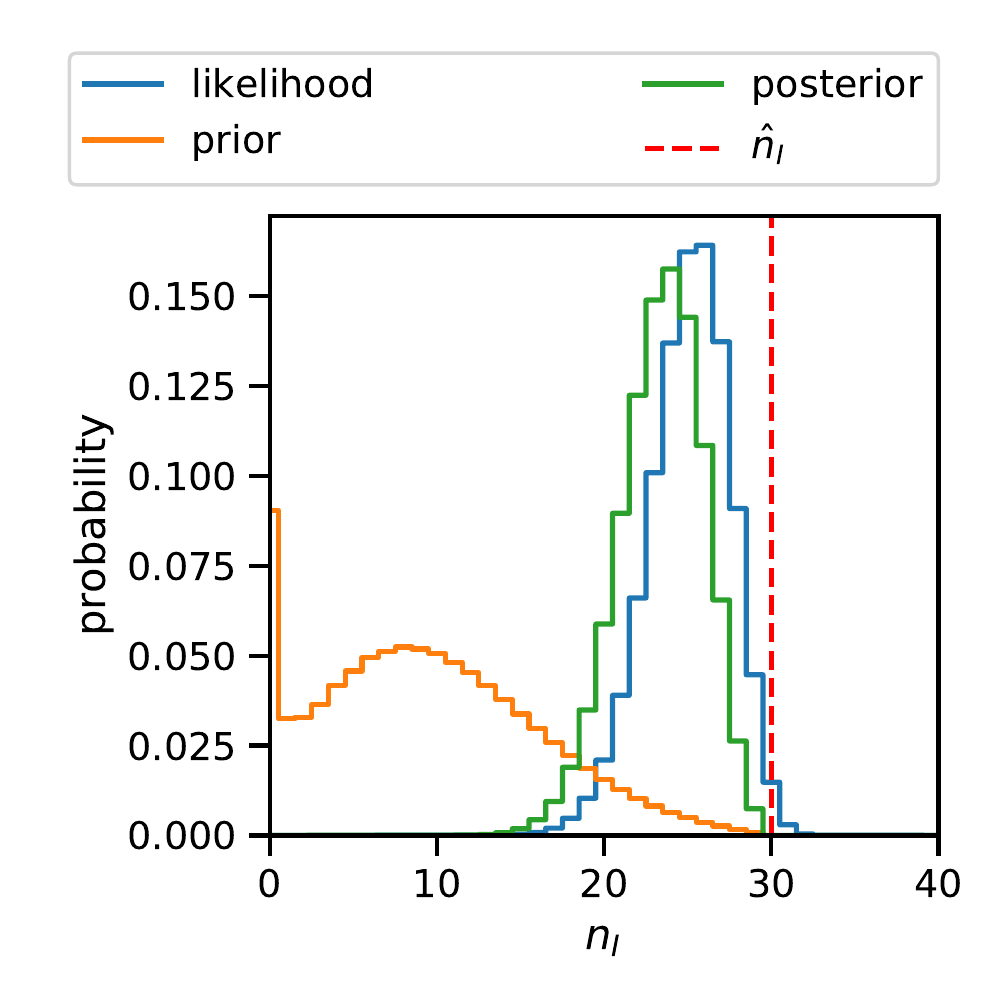}
    \end{minipage}
    \begin{minipage}{0.24\textwidth}
    \centering
    \includegraphics[scale=.4]{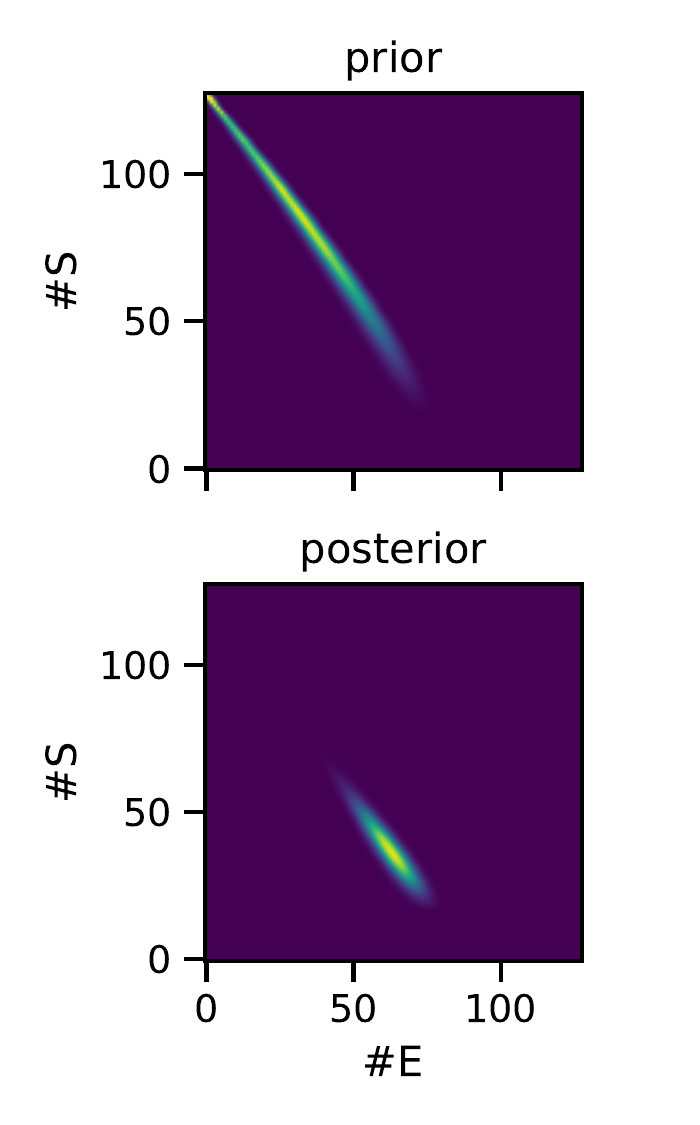}
    \end{minipage}
    \caption{(left) A comparison of the prior dynamics and the posterior smoothing (bridging) dynamics. (middle) The prior, likelihood, and posterior of the number of infected individuals $n_I$ at time $t=0.3$ given the measurement $\hat{n}_I=30$. (right) The prior and posterior distribution over the latent types $E$ and $S$.}
    \label{fig:seir}
\end{figure}

% The first sub-problem is the estimation of the posterior distribution at time $t$ given the measurement.
The posterior for $n_I$ infected individuals at time $t$, given measurement $Y_t=\hat{n}_I$ can be computed using  Bayes' rule
\begin{equation}\label{eq:bayes_posterior}
\Pr(X_t^{(I)}=n_I\mid Y_t=\hat{n}_I)\propto \Pr(Y_t=\hat{n}_I\mid X_t^{(I)} = n_I)\Pr(X_t^{(I)}=n_I)\,.
\end{equation}
This problem is an extension of the bridging problem discussed up until now.
The difference is that the terminal posterior   is   estimated it using the result of the lumped forward equation and the measurement distribution using~\eqref{eq:bayes_posterior}.
Based on this estimated terminal posterior, we compute the bridging probabilities and refine the truncation tailored to the location of the posterior distribution.
In Figure~\ref{fig:seir} (left), we illustrate the bridging distribution between the terminal posterior and initial distribution.
In the context of filtering problems this is commonly referred to as smoothing.
Using the learned truncation, we can obtain the posterior distribution for the number of infected individuals at $t=0.3$ (Figure~\ref{fig:seir} (middle)).
Moreover, can we infer a distribution over the unknown number of susceptible and exposed individuals (Figure~\ref{fig:seir} (right)).

% Using the posterior \eqref{eq:bayes_posterior} of the first sub-problem, we can proceed and analyze the bridging problem from using the terminal constraint of the infected population exceeding threshold $k$.
% To this end this area of the state-space is made absorbing and we can proceed as before.
\section{Conclusion}
The analysis of Markov Jump processes with constraints on the initial and terminal behavior is an important part of many probabilistic inference tasks such as parameter estimation using Bayesian or maximum likelihood estimation, inference of latent system behavior, the estimation of rare event probabilities, and reachability analysis for the verification of temporal properties.  
If endpoint constraints correspond to atypical system behaviors, 
standard analysis methods fail
as they have no strategy to identify those parts of the state-space relevant for meeting the terminal constraint.

Here, we proposed a method that is not based on stochastic sampling and statistical estimation 
but provides a direct numerical approach.
It starts with an abstract lumped model, which is iteratively refined such that only those parts of the model are considered that contribute to the  
probabilities of interest.
In the final step of the iteration, we operate at the granularity of the original model and compute lower bounds for these bridging probabilities that are rigorous up to the error of the numerical integration scheme.
%To achieve accurate bounds, appropriate error tolerances for the numerical integration have to be chosen.

Our method exploits the population structure of the model, which is present in many important application fields of MJPs.
% As with any direct method based on truncations, our approach is limited to models where population sizes remain relatively small.
Based on experience with other work based on truncation, the approach can be expected to scale up to at least a few million states~\cite{mikeev2011efficient}.
Compared to previous work, our method neither relies on approximations of unknown accuracy nor additional information such as a suitable change of measure in the case of importance sampling.
It only requires a truncation threshold and an initial choice for the macro-state sizes.
% Alternatively, uniformization can be applied to strengthen the precision guarantees \cite{mateescu2010fast}.\MB{for backward solutions too?; uniformization does not fit with the paragraphs opening; schon oben}

In future work, we plan to extend our method to hybrid approaches, in
which a moment representation is employed for large populations while discrete counts are maintained for small populations.
Moreover, we will apply our method to model checking  where constraints are described by some temporal logic \cite{hajnal2019data}.
\subsubsection{Acknowledgements}
This project was supported by the DFG project MULTIMODE and Italian PRIN project SEDUCE.

\bibliographystyle{splncs04}
\bibliography{paper.bib}

% \vfill

% \vfill

% {\small\medskip\noindent{\bf Open Access} This chapter is licensed under the terms of the Creative Commons\break Attribution 4.0 International License (\url{http://creativecommons.org/licenses/by/4.0/}), which permits use, sharing, adaptation, distribution and reproduction in any medium or format, as long as you give appropriate credit to the original author(s) and the source, provide a link to the Creative Commons license and indicate if changes were made.}

% {\small \spaceskip .28em plus .1em minus .1em The images or other third party material in this chapter are included in the chapter's Creative Commons license, unless indicated otherwise in a credit line to the material.~If material is not included in the chapter's Creative Commons license and your intended\break use is not permitted by statutory regulation or exceeds the permitted use, you will need to obtain permission directly from the copyright holder.}

% \medskip\noindent\includegraphics{cc_by_4-0.eps}

\end{document}